\documentclass[conference]{IEEEtran}
\IEEEoverridecommandlockouts
\usepackage{cite}
\usepackage{amsmath,amssymb,amsfonts}
\usepackage{algorithmic}
\usepackage{algorithm} 
\usepackage{graphicx}
\usepackage{textcomp}
\usepackage{xcolor}
\usepackage{multirow}
\usepackage{subfig}
\usepackage[compatibility=false]{caption}
\def\BibTeX{{\rm B\kern-.05em{\sc i\kern-.025em b}\kern-.08em
    T\kern-.1667em\lower.7ex\hbox{E}\kern-.125emX}}
\begin{document}

\title{SpeechGuard: Online Defense against Backdoor Attacks on Speech Recognition Models\\
}

\author{\IEEEauthorblockN{1\textsuperscript{st} Jinwen Xin}
\IEEEauthorblockA{\textit{School of Cyber Engineering} \\
\textit{Xidian University}\\
Xi'an, China \\
904175502@qq.com}
\and
\IEEEauthorblockN{2\textsuperscript{nd} Xixiang lv*}
\IEEEauthorblockA{\textit{School of Cyber Engineering} \\
\textit{Xidian University}\\
Xi'an, China \\
xxlv@mail.xidian.edu.cn}
\thanks{\scriptsize Author's accepted manuscript of the IEEE IJCNN 2024 paper; not the IEEE Version of Record. DOI: 10.1109/IJCNN60899.2024.10650300.}
\thanks{\scriptsize \textcopyright 2024 IEEE. Personal use of this material is permitted. Permission from IEEE must be obtained for all other uses, in any current or future media, including reprinting/republishing this material for advertising or promotional purposes, creating new collective works, for resale or redistribution to servers or lists, or reuse of any copyrighted component of this work in other works.}
}
\maketitle
\begin{abstract}
Backdoor attacks pose a critical threat to neural network models, allowing attackers to implant a backdoor during the training phase by manipulating a small portion of the training data. In security-sensitive applications such as voice interaction for autonomous driving, the presence of backdoor attacks introduces substantial security risks. This study focuses on implementing backdoor defense measures for speech recognition models in run-time, taking into account the characteristics of audio signals. We propose SpeechGuard, the first online backdoor defense pipeline designed to identify and purify poisoned audio samples. Specifically, we improve STRIP method to perform adaptive perturbation injection to detect and filter poisoned samples, named as S-STRIP. More importantly, we further consider the purification of poisoned samples. We utilize time-frequency (T-F) masking to suppress the expression of trigger signals and autonomously generate masks based on an autoencoder. The two-stage processing prevents the backdoor in the model from being triggered, and even input speech carrying triggers can be accurately predicted. Extensive experimental demonstrate that SpeechGuard can accurately filter out poisoned samples. Through purification, it can significantly mitigate the backdoor threat while maintaining a certain prediction accuracy.
\end{abstract}

\begin{IEEEkeywords}
Neural Networks, Speech Recognition, Backdoor Attack, Backdoor Defense
\end{IEEEkeywords}

\section{Introduction}
As a breakthrough technology in the field of artificial intelligence, deep learning has achieved remarkable results in image recognition, machine translation, speech recognition and other fields, driving the progress of technology and society. Neural networks (NN), as the core of deep learning, have been proven to be vulnerable to various attacks due to their inherent black-box nature and high data dependency. For instance, adversarial example attacks can fool autonomous vehicles by adding imperceptible perturbations to the traffic sign\cite{li2020adaptive}, while data poisoning attacks may inject carefully crafted comments to maliciously manipulate a recommender system\cite{Huang2021data}.
\par
Recently, another attack threat against NN has caught the attention of researchers, known as \emph{backdoor attacks}. In general, backdoor attacks work by implanting a backdoor in models during the training stage so that the infected models perform well on benign samples, whereas their predictions will be maliciously altered if the buried backdoor is activated by the preset trigger\cite{li2022backdoor}. Early research on backdoor attacks focused on the image and text domains, but recent findings have shown that backdoor attacks have gradually penetrated into the \emph{speech recognition} domain\cite{koffas2022can,liu2022opportunistic,zong2022trojanmodel}. The adversary can successfully activate the malicious behavior of the victim model by using an ultrasonic pulse as the inaudible trigger\cite{koffas2022can}, and the trigger can also be ambient noise where the backdoor can be passively triggered\cite{liu2022opportunistic}. In real-world physical scenarios (e.g., voice assistant), speech recognition systems may face disturbances from various environmental noises that could potentially contain malicious triggers. Therefore, it is imperative to explore corresponding defense measures to address this emerging security challenge.
\par
We focus on \emph{online} backdoor defense measures in speech recognition tasks. Advanced defense schemes are typically evaluated on image classification tasks\cite{gao2019strip,li2021anti}, which is fundamentally different from speech recognition tasks in the realm of input space. Therefore, it is crucial to design effective detection methods that consider the specific characteristics of audio signals. Furthermore, prevalent online defense strategies primarily concentrate on detecting poisoned samples but lack mechanisms for subsequent processing. In real-time speech recognition systems, a straightforward solution is to directly reject poisoned voice inputs, which effectively deals with malicious visitors but may degrade the confidence of legitimate users. We underscore the importance of purifying poisoned samples and ensuring the completion of the inference process.
\par
In this paper, we propose SpeechGuard, an online backdoor defense scheme designed specifically for speech recognition tasks. SpeechGuard introduces a two-stage defense pipeline that seamlessly integrates poisoned sample \emph{detection} and \emph{purification}. Specifically, given an infected model that has been deployed and running, the input voice undergoes a two-stage processing to mitigate the backdoor threat. In the detection stage, SpeechGuard employs S-STRIP, an enhanced version of STRIP\cite{gao2019strip} achieved through improved \emph{perturbation injection}, to identify and filter poisoned samples. The filtered samples then proceed to the second stage. In the purification stage, SpeechGuard trains an \emph{autoencoder} to learn the mapping from poisoned inputs to the time-frequency mask.  This mask is then utilized to suppress trigger signals for the purification of poisoned samples. The effectiveness of this purification method stems from the \emph{sparsity} of speech signals and the \emph{isolation} of trigger signals in the T-F domain. The combination of two stages is effective in identifying and purifying poisoned samples. Ensuring a certain level of prediction accuracy for the purified poisoned samples, it effectively mitigates the risk associated with backdoors.
\par
Our contributions can be summarized as follows:
\begin{itemize}
  \item We improved the perturbation injection strategy in STRIP method, making it more suitable for detecting poisoned audio inputs.
  \item We focus on the purification of poisoned audio inputs. We utilize T-F masking to suppress trigger signals and generate masks based on an autoencoder.
  \item We propose SpeechGuard, an online backdoor defense pipeline for speech recognition models, which provides a two-stage backdoor defense including poisoned sample detection and purification.
  \item We conduct extensive experiments to evaluate the performance of SpeechGuard. Experimental results indicate that SpeechGuard can accurately detect poisoned samples and significantly mitigate the backdoor threat introduced by trigger signals.
\end{itemize}

\section{Background}

\subsection{Related Work}
\subsubsection{Backdoor Attack} 
Gu et al. \cite{gu2019badnets} first proposed the \emph{BadNets} attack scheme based on data poisoning for outsourced training and transfer learning scenarios. Liu et al. \cite{liu2018trojaning} generated the trojan trigger by reversing neurons, followed by retraining the model with external data to inject the backdoor. Chen et al. \cite{chen2017targeted} believed that, in advanced backdoor attack schemes, poisoned samples should be indistinguishable from clean samples to evade manual detection. They proposed the \emph{blended injection} strategy, which can make triggers more invisible by reducing the blend ratio. Subsequently, researchers worked on designing more stealthy and complicated backdoor triggers. Liu et al. \cite{liu2020reflection} proposed a novel attack scheme that used \emph{reflection}, a common natural phenomenon, as a trigger, which was highly stealthy and maintained a high attack success rate. Cheng et al. \cite{cheng2021deep} adopted \emph{style transfer} to implement backdoor attacks in feature space. Unlike injecting triggers in pixel space, the activation of backdoor relied on the deep-level features of poisoned samples.
\subsubsection{Backdoor Defense}
Various studies have proposed defense schemes against the threat of backdoor attacks. Liu et al. \cite{liu2018fine} pruned the neurons associated with the trigger to eliminate the backdoor. The pruned model was then fine-tuned with clean samples to restore prediction accuracy. Wang et al. \cite{wang2019neural} proposed the first trigger synthesis-based backdoor detection method, called \emph{Neural Cleanse}. Li et al. \cite{li2021anti} proposed a general defense scheme---\emph{Anti-Backdoor Learning}, which could automatically prevent backdoor attacks during the training process. The above defense schemes based on \emph{model reconstruction} or \emph{trigger synthesis} aim at detecting and mitigating the backdoor before deployment, also known as \emph{offline defense}. \emph{Online defense} aims at detecting poisoned inputs and eliminating the impact of backdoor during the \emph{run-time} phase. Doan et al. \cite{doan2020februus} proposed a preprocessing-based backdoor defense scheme \emph{Februus}, which adopted GradCAM \cite{selvaraju2017grad} to identify suspected trigger regions and remove them by surgery. Gao et al. \cite{gao2019strip} proposed STRIP, a perturbation-based \emph{run-time} poisoned sample detection scheme. In the inference stage, poisoned samples were detected based on the randomness of the predicted outcome of perturbed inputs.
\subsubsection{Research Focused on Speech Recognition} Recently, attack schemes focused on speech recognition tasks were proposed. Koffas et al. \cite{koffas2022can} adopted an ultrasonic pulse as a trigger to implement the inaudible backdoor attack. Liu et al. \cite{liu2022opportunistic} explored the first audible backdoor attack paradigm for speech recognition, characterized by passively triggering and opportunistically invoking. Zhai et al. \cite{zhai2021backdoor} designed a clustering-based attack scheme to implement backdoor attacks on speaker verification models. There is currently a gap in backdoor defense solutions focused on speech recognition systems. While defense schemes suitable for image and text domains may be equally applicable to speech recognition tasks  \cite{gao2019strip,li2021anti}, further validation of their effectiveness is needed. Additionally, given the specificity of speech signals, further exploration of defense pipeline for speech recognition models is then needed.

\subsection{Mathematical Description of Backdoor Attacks on Speech Recognition Tasks}
\subsubsection{Training procedure of speech recognition models}
In general, let $\mathcal{D}_{train}=\{(x_i,y_i)\}_{i=1}^N$ represent the original audio dataset with $N$ clean audio samples, where ${x_i}\in{X}$ denotes the time series representation of the sample, and ${y_i}\in{Y}=\{1,2,\cdots ,K\} $ signifies the true label of the input $x_i$. For a speech command recognition task, the objective is to learn a model $\mathcal{F}_{\theta} :X\rightarrow{Y}$, where $X$ denotes the input space and $Y$ denotes the label space. The goal of model training is to find the optimal parameter $\theta$ to minimize the distance between the prediction results of the model $\mathcal{F}_{\theta}$ and the true labels, typically expressed through the loss function $\mathcal{L}$. The parameters are optimized during training as \eqref{eq:1}:
\begin{equation}
\theta ^*=\mathop{\arg\min}\limits_{\theta}\sum_{i=1}^N \mathcal{L}(\mathcal{F}_{\theta} (x_i),y_i).
\label{eq:1}
\end{equation}
\subsubsection{Backdoor attacks based on data poisoning}In a data poisoning-based backdoor attack, the attacker typically generates poisoned samples by modifying a small subset of clean samples within the original training set: $\mathcal{D}_{poison}=\{(G_{\delta}(x_i),t)\}_{i=1}^P$, where $G_{\delta}:X\rightarrow X$ represents the attacker-designed method for generating poisoned samples using trigger signals $\delta$, $t$ denotes the attacker's pre-defined target label. Followed by incorporating the poisoned samples to create the poisoning training set: $\mathcal{D}_{train}^*=\mathcal{D}_{train}\bigcup \mathcal{D}_{poison}$ . Training on the poisoned dataset will result in the backdoor model $\mathcal{F}_{ \theta} ^*$. For the test set $\mathcal{D}_{test}=\{(x_i,y_i)\}_{i=1}^M$, the backdoor model will maintain prediction accuracy on benign inputs: $\mathcal{F}_{\theta}^*(x_t)=y_t$, whereas for inputs containing trigger signals, the predictions will be directed towards the target label: $\mathcal{F}_{\theta}^*(G_{\delta}(x_t))=t $. The poisoning rate is defined as $\frac{P}{N}$.

\section{Threat Model and Defense Goals}
\subsection{Attack Scenario}
The training of speech recognition models depends on extensive audio data and significant computing resources. Confronted with elevated training costs, users frequently seek assistance from service providers. These services include adopting third-party datasets, outsourcing training and employing third-party pre-trained models\cite{li2022backdoor}. However, models trained using the aforementioned services are susceptible to backdoor attacks. Given that SpeechGuard is deployed in run-time phase, it is effective for all the aforementioned attack scenarios. 
\subsection{Capacity Limitation}
The capabilities and knowledge of the attacker and defender are delineated as follows:
 \begin{itemize}
  \item \textbf{Attacker}: We assume that the attacker implements backdoor attacks based on data poisoning. The attacker can access raw training data and make modifications, has knowledge of the model architecture and parameters, and can control the training procedure. In addition, the attacker can arbitrarily choose the type, duration and insertion position of the trigger, while triggers are \emph{sample-agnostic}. The aforementioned assumptions are generally followed in attack scenarios.
  \item \textbf{Defender}: Defense is extremely simple and does not require additional conditions, except for a few clean samples. The defender deploys defense during the run-time of the target model. The defender cannot obtain the architecture and parameters of the victim model, and has no control over the training procedure. Meanwhile, the defender can obtain a portion of the clean samples, but may not have access to the poisoned samples. This assumption is reasonable as the attacker will not disclose poisoned samples which are indistinguishable from clean samples, while users typically retain a few clean samples for verifying model performance.
\end{itemize}
\subsection{Defense Goals}
SpeechGuard endeavors to institute an effective defense during the run-time phase, with three articulated goals:
\begin{itemize}
  \item \textbf{Filtering out poisoned samples.} SpeechGuard can identify potentially poisoned voice inputs carrying triggers.
  \item \textbf{Eliminate the impact of trigger.} SpeechGuard can eliminate trigger signals in the filtered poisoned samples and then they will be accurately predicted.
  \item \textbf{Maintain prediction accuracy.} The defense does not cause a notable loss in prediction accuracy of clean samples.
\end{itemize}

\section{SpeechGuard}
\subsection{System Overview}
Existing backdoor attack schemes predominantly follow the trigger-backdoor activation mechanism, which establishes a robust connection between trigger features and backdoor behavior. Disrupting either component renders the attack ineffective. During the inference phase, SpeechGuard disrupts the trigger-backdoor by eliminating the trigger embedded in poisoned samples. It provides a two-stage protection pipeline including poisoned sample detection and poisoned sample purification. Fig.~\ref{fig:1} shows an overview of SpeechGuard, which consists of the following two stages:
\begin{figure*}[htbp]
  \centering
  \includegraphics[width=0.8\textwidth]{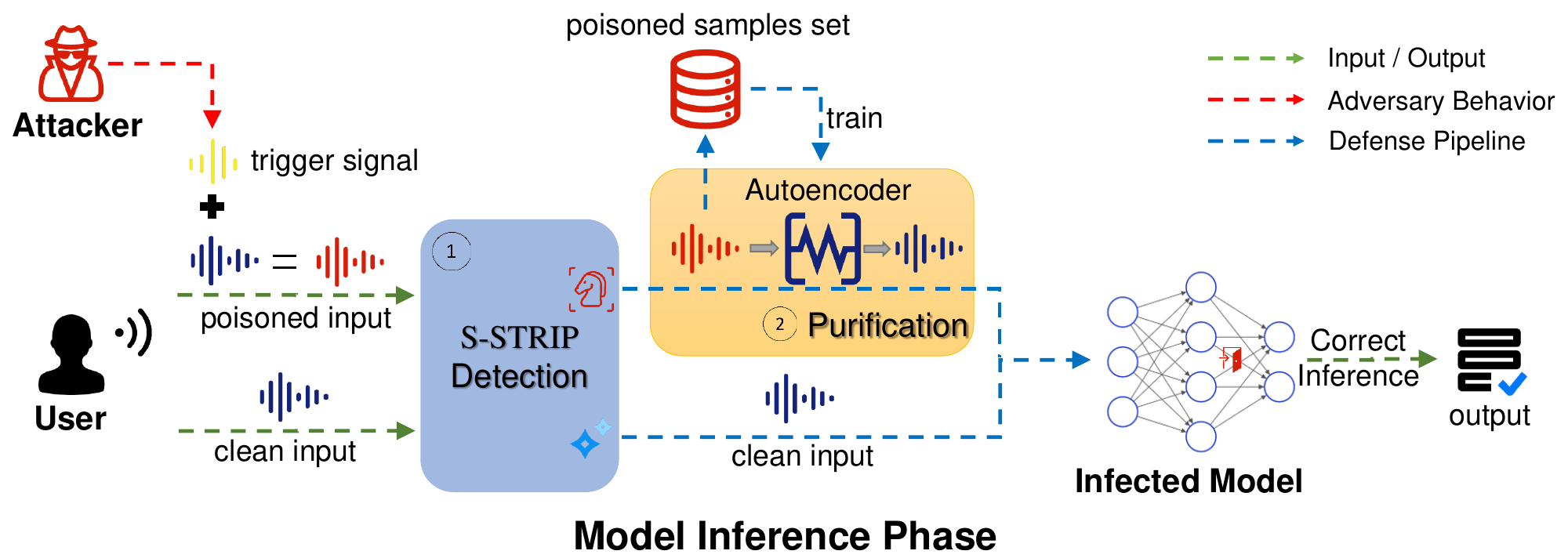} % 替换为你的图片文件名
  \caption{An overview of SpeechGuard}
  \label{fig:1}
\end{figure*}
\subsubsection{Poisoned Sample Detection} 
At this stage, we use an improved STRIP \cite{gao2019strip} to detect poisoned inputs. Specifically, the initial STRIP adds strong perturbations to the input samples by directly superimposing perturbations and without considering the input signal strength, which results in an uneven energy distribution of the perturbed inputs. In view of this problem, we propose adding perturbations based on a predetermined signal-to-noise ratio. We name the improved method as S-STRIP, making it more applicable to speech recognition tasks.
\subsubsection{Poisoned Sample Purification} 
At the second stage, based on the poisoned samples filtered in the previous stage, SpeechGuard trains an autoencoder to learn the mapping from the poisoned input signals to the T-F mask. The decoding part of the autoencoder is used as a generator to generate the masks to suppress the backdoor trigger signals in the poisoned inputs, thereby achieving the purification of poisoned samples. The feasibility of this purification method stems from leveraging the sparse distribution of audio signals in the time-frequency domain.

\par
The rest of this section will dive into the technical details of SpeechGuard.
% \begin{algorithm}[t]
% \caption{Generation of Perturbation Samples} 
% \label{alg1} 
% \begin{algorithmic}[1] 
% \REQUIRE Variables: $x$, $p$, $l_1$, $l_2$, $\alpha$
% \ENSURE Variables: $x^p$ 
% % for loop
% \FOR{$i=1$ to $l_1$}
% \IF{$i\leq l_2$}
% \STATE $x_i^p \leftarrow x_i + \alpha \cdot p _i$
% \ELSE
% \STATE $x_i^p\leftarrow x_i$
% \ENDIF
% \ENDFOR
% \end{algorithmic}
% \end{algorithm}

\subsection{Poisoned Sample Detection Based on S-STRIP}
\subsubsection{A Brief Overview of STRIP}
The STRIP's detection capability relies on a fundamental observation: When multiple strong perturbations are added to a poisoned sample, the prediction results of all perturbed inputs tend to converge towards the target label. In contrast, clean samples undergoing the same robust perturbations exhibit substantial alterations in the prediction results. The output alterations can be measured by information entropy. Consequently, intentionally applying robust perturbations to input samples allows for discerning whether the input is poisoned. Specifically, input $x$ is replicated  to generate $N$ input replicas. Subsequently, multiple strong perturbations $p_i$ are applied to each replica, resulting in a perturbed input set $x^{p_n}\in\{x^{p_1},x^{p_2},\cdots ,x^{p_N}\}$. The Shannon entropy for all N perturbed inputs is computed according to \eqref{eq:2}:
\begin{equation}
\mathbb{H}=\frac{1}{N}\sum_{n=1}^{N}\left(-\sum_{i=1}^{M}y_{n,i}\times\log_2(y_{n,i})\right).
\label{eq:2}
\end{equation}
Here, $y_{n,i}$ represents the probability of the $n$-th perturbed sample belonging to class $i$, and $M$ is the total number of classes in the training task. The Shannon entropy $\mathbb{H}$ serves as a metric for determining whether the input sample $x$ is poisoned. If $\mathbb{H}$ falls below the predefined threshold $T$ (estimated based on the entropy distribution of benign samples), the input $x$ is classified as a poisoned sample.
\subsubsection{Perturbation Method in S-STRIP}
Similar to introducing perturbations in the pixel space of images, we opt to add perturbations into the time series of audio inputs. These perturbations are derived either from the dataset or random noise. Specifically, the temporal waveform representation of the raw audio input is denoted as $x=\{a_1,a_2,a_3\cdots a_{l_1}\}$, while the perturbation signal is represented as $p=\{b_1,b_2,b_3\cdots b_{l_2}\}$. The generation process of the perturbed one $x^p$ is described in \eqref{eq:3}: 
\begin{equation}
x^p = x + \alpha \cdot p,
\label{eq:3}
\end{equation}
symbol $\alpha$ represents the blend ratio. Adjusting the value of $\alpha$ allows for adding perturbations with a certain Signal-to-Noise Ratio (SNR). And SNR is expressed as \eqref{eq:4}:
\begin{equation}
SNR(S(t),N(t))=10\log_{10}{\left ( \frac{\sum_{t}S^2(t)}{\sum_{t}N^2(t)} \right )},
\label{eq:4}
\end{equation}
where $N(t)$ and $S(t)$ represent the noise signal and the clean signal, respectively. When SNR is set to $q~dB$, the blend ratio $\alpha$ can be calculated with \eqref{eq:5}:
\begin{equation}
\alpha=\sqrt{\frac{\sum_{t}x^2(t)}{10^{\frac{q}{10}}\sum_{t}p^2(t)}}.
\label{eq:5}
\end{equation}
Following the aforementioned method, the defender can introduce perturbations with appropriate intensity.

\subsection{Poisoned Sample Purification Based on T-F Masking}
\subsubsection{The Principle of suppressing trigger signals with the T-F Masking} 
We briefly introduce the concept of time-frequency masking and analyze the feasibility of utilizing masks to eliminate trigger signals in the time-frequency domain.
\paragraph{Time-frequency masking} A T-F mask is a binary or soft mask that is applied to a T-F representation of an audio sample, such as a spectrogram obtained through the Short-Time Fourier Transform (STFT). The STFT calculation process is defined in \eqref{eq:6}:
\begin{equation}
X(t,f)=\int_{-\infty}^{\infty} w(t-\tau)x(\tau)e^{-j 2 \pi f\tau }\, d\tau ,
\label{eq:6}
\end{equation}
where $X(t,f)$ represents a two-dimensional matrix, signifying the energy of the signal $x(t)$ at the $t$-th time frame and the $f$-th frequency band, and ${w}(t)$ denotes the window function. Assuming the existence of an ideal T-F mask $M(t,f)$, wherein regions corresponding to valid speech signals are assigned a value of 1, and the remaining units (including trigger signals) are assigned a value of 0. T-F Masking is to mutiply the mask $M(t,f)$ with the poisoned input signals $Y(t,f)$, which yields the clean speech signals $\widehat{X}(t,f)$, expressed as \eqref{eq:7}: 
\begin{equation}
\widehat{X}(t,f)=\widehat{M}(t,f)\times Y(t,f).
\label{eq:7}
\end{equation}
This aforementioned masking is known as the Ideal Binary Mask (IBM),  which can be computed based on the relative magnitudes of the original signal energy and the trigger signal energy. IBM is calculated as shown in \eqref{eq:8}:
\begin{equation}
\mathbf{M}_{IBM}(t,f) =
\begin{cases} 
\text{1},  & \text{if } |X(t,f)|^2-|N(t,f)|^2>0 \\
\text{0}, & \text{otherwise }.
\end{cases}
\label{eq:8}
\end{equation}
\paragraph{Time-frequency analysis}
The adoption of T-F masking for purification is based on the following intuitions: (i) The distribution of valid speech signals exhibits sparse characteristics. (ii) Audio signals acting as triggers can be regarded as a form of noise, typically not residing in the same frequency bands as speech signals. (iii) To make attacks more covert and threatening, attackers prefer to embed triggers in frequency bands isolated from the speech signal (e.g., ultrasound \cite{koffas2022can}). Assuming the clean input signal $x(t)$ is convolved with the trigger signal $\delta(t)$ in the time domain, resulting in the poisoned input signal $y(t)$. It is challenging to separate the clean speech signal $x(t)$ from poisoned input signal $y(t)$ in the time domain. Therefore, we endeavor to purify the poisoned input signal in the time-frequency domain. By analyzing the spectrogram of signals (Figure 2), it can be observed that the audio signal exhibits a characteristic of sparse distribution. Additionally, there is relatively less overlap between the clean input signal (green region) and the trigger signal (red region) in the time-frequency domain. Therefore, it is feasible to purify the poisoned input signal in the time-frequency domain based on this non-overlapping characteristic.
\begin{figure}
\centering
\includegraphics[width=250pt]{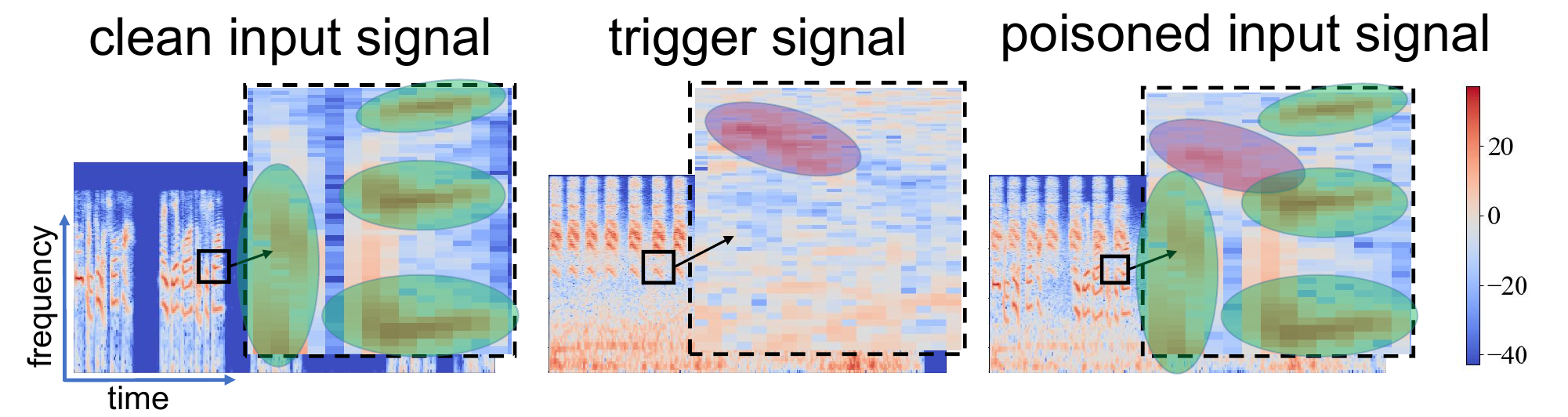}
\caption{The spectrogram of the audio signal.}
\label{fig:2}
\end{figure}

\subsubsection{Generating masks using an autoencoder} 
\begin{figure}
\centering
\includegraphics[width=250pt]{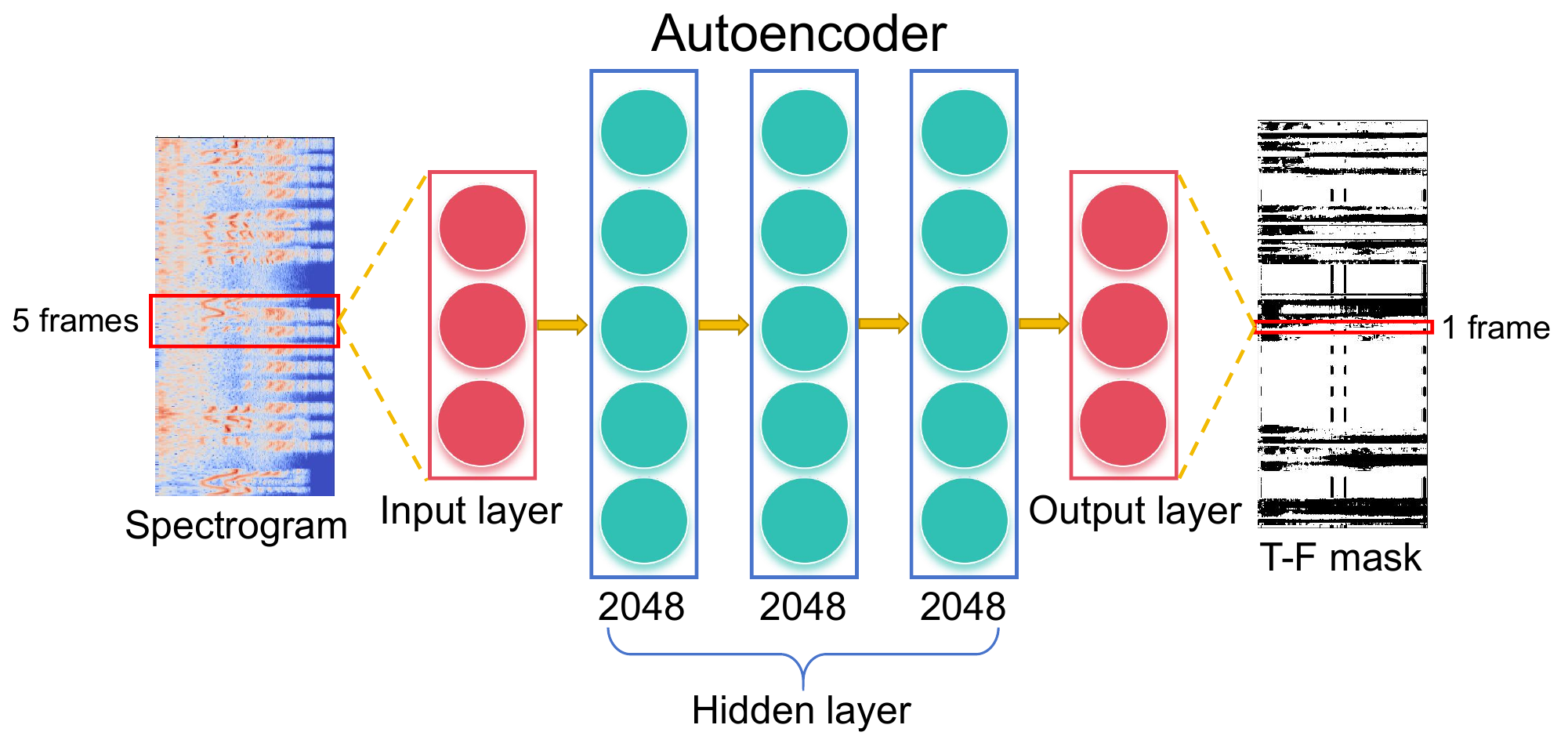}
\caption{Autoencoder architecture.}
\label{fig:3}
\end{figure}

As an unsupervised model, autoencoder is used to compress and encode input data, and then decode and reconstruct the original data. In this study, we propose an autoencoder architecture (Fig.~\ref{fig:3}) designed to map poisoned input signals to T-F masks.
\paragraph{Autoencoder architecture} We employ a fully connected neural network with three hidden layers as an autoencoder, where each hidden layer consists of 2048 neurons. We utilize LeakyReLU as the activation function, while incorporating Dropout and Regularization to prevent overfitting. Given that the target output values lie within the range of $[0, 1]$, we utilize the Sigmoid activation function for the final layer.
\paragraph{Autoencoder training} The poisoned input samples detected by S-STRIP are transformed into T-F representations and then serve as inputs to the autoencoder. Considering that effective speech information is transmitted across time frames, adjacent 5 frames are concatenated into a feature vector through frame expansion. The labels are IBM masks, computed based on the SNR between the poisoned and clean speech signals. Employing Mean Squared Error as the loss function and Adaptive Gradient Descent as the optimizer, the autoencoder aims to minimize the discrepancy between the predicted mask and the actual mask, as defined in \eqref{eq:9}
\begin{equation}
\theta ^*=\mathop{\arg\min}\limits_{\theta}\sum_{i=1}^N || M_i - f_{\theta}(X_i) ||_2^2.
\label{eq:9}
\end{equation}
Here, $X_i$ represents the T-F coefficients of the $i$-th training sample, $M_i$ represents the target masking, and $\theta$ indicates the training parameters of autoencoder $f$.
\paragraph{Purifying poisoned samples using the autoencoder}
During the inference stage, the filtered poisoned speech input $y_t$ is transformed into T-F representation $Y_t$. Subsequently, $Y_t$ is fed into the pre-trained autoencoder to generate the mask $M_t$. The poisoned input signal $Y_t$ is then subjected to element-wise multiplication with the mask $M_t$ to derive the magnitude spectrum of the clean input signal $X_t$, as depicted in \eqref{eq:10}:
\begin{equation}
M_t = f_{\theta}(Y_t),\ X_t = Y_t \odot M_t,
\label{eq:10}
\end{equation}
where $\odot$ denotes the Hadamard product. This process will mask the T-F units belonging to trigger signals while preserving the T-F units associated with valid speech signals. Finally, the magnitude spectrum and phase spectrum are integrated to reconstruct the signal. The Inverse Short-Time Fourier Transform (ISTFT) is subsequently applied to convert it back into the time-domain signal $x(t)$. The computation process for the $n$-th frame signal $x_n(t)$ is delineated as \eqref{eq:11}:
\begin{equation}
x_n(t)=\mathcal{F}^{-1}\left \{ X_n(f)e^{j\phi_n(f) } \right \},
\label{eq:11}
\end{equation}
where $\mathcal{F}^{-1}$ represents the ISTFT, and $\phi_n(f)$ denotes the phase spectrum of the $n$-th frame. The purified speech input can then be further fed into the speech recognition model to perform subsequent inference tasks. Through the above detection and purification pipeline processing, even if the model is implanted with a backdoor, the activation of the backdoor behavior will fail due to the weakening of the trigger.

\section{Experimental Evaluation}
We conduct experiments on two baseline datasets and models to assess the performance of SpeechGuard in countering mainstream poisoning attack schemes.
\subsection{Experiment Setup}
\subsubsection{Datasets and Model Architectures}
We select two popular audio datasets typically employed for keyword recognition tasks as our training task. Below is a brief description:
\begin{itemize}
  \item \textbf{SCDv2}: Speech Commands Dataset Version 2 (SCDv2)\cite{warden2018speech} comprises 30 different commands extracted from human speech segments. We have selected 10 commands to form a 10-class speech recognition task.
  \item \textbf{AMT}: AudioMNIST (AMT) \cite{becker2018interpreting} serves as a resource for speech recognition, featuring audio segments representing numbers from 0 to 9. These audio segments are derived from the corresponding digits in the MNIST dataset.
\end{itemize}
We choose two widely recognized neural networks as victim models and provide concise introductions for each:
\begin{itemize}
  \item \textbf{2D-CNN}: The 2D-CNN model is a lightweight convolutional neural network commonly employed for keyword detection\cite{samizade2020adversarial}. It consists of three convolutional layers and three fully connected layers.
  \item \textbf{Att-LSTM}: The Att-LSTM model is a convolutional recurrent neural network with the attention mechanism \cite{de2018neural}. It serves as a lightweight model that can be deployed on mobile devices and executed locally.
\end{itemize}
\subsubsection{Attack Configuration}
Backdoor attacks on speech recognition models primarily employ data poisoning method. The triggers can be categorized into three types, each illustrated in the spectrogram depicted in Fig.~\ref{fig:4}:
\begin{figure}
\centering
\subfloat[Random noise]{\includegraphics[width=.32\linewidth,height=50pt]{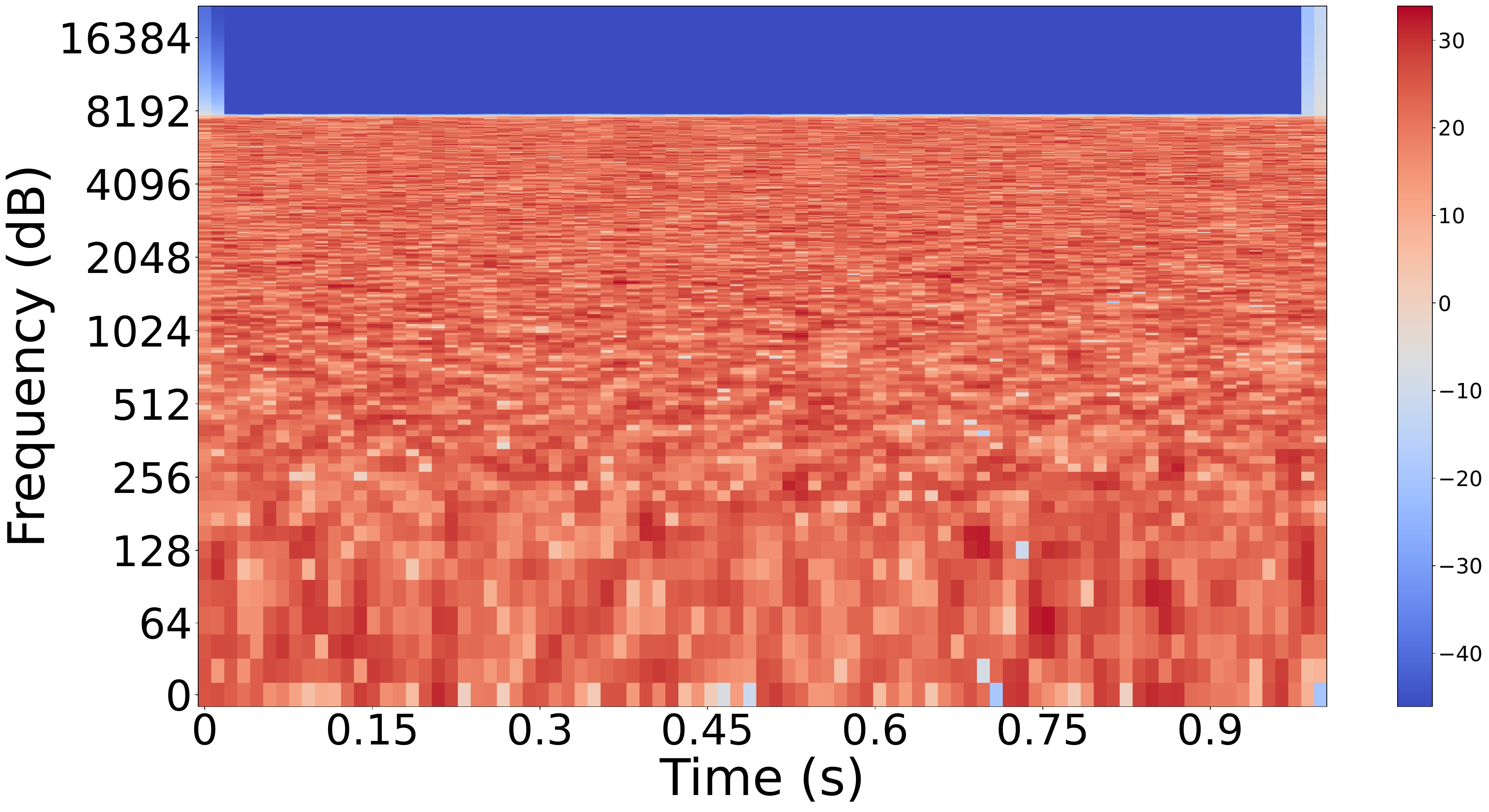}%
\label{Random noise}}
\hfil
\subfloat[Environmental noise]{\includegraphics[width=.32\linewidth,height=50pt]{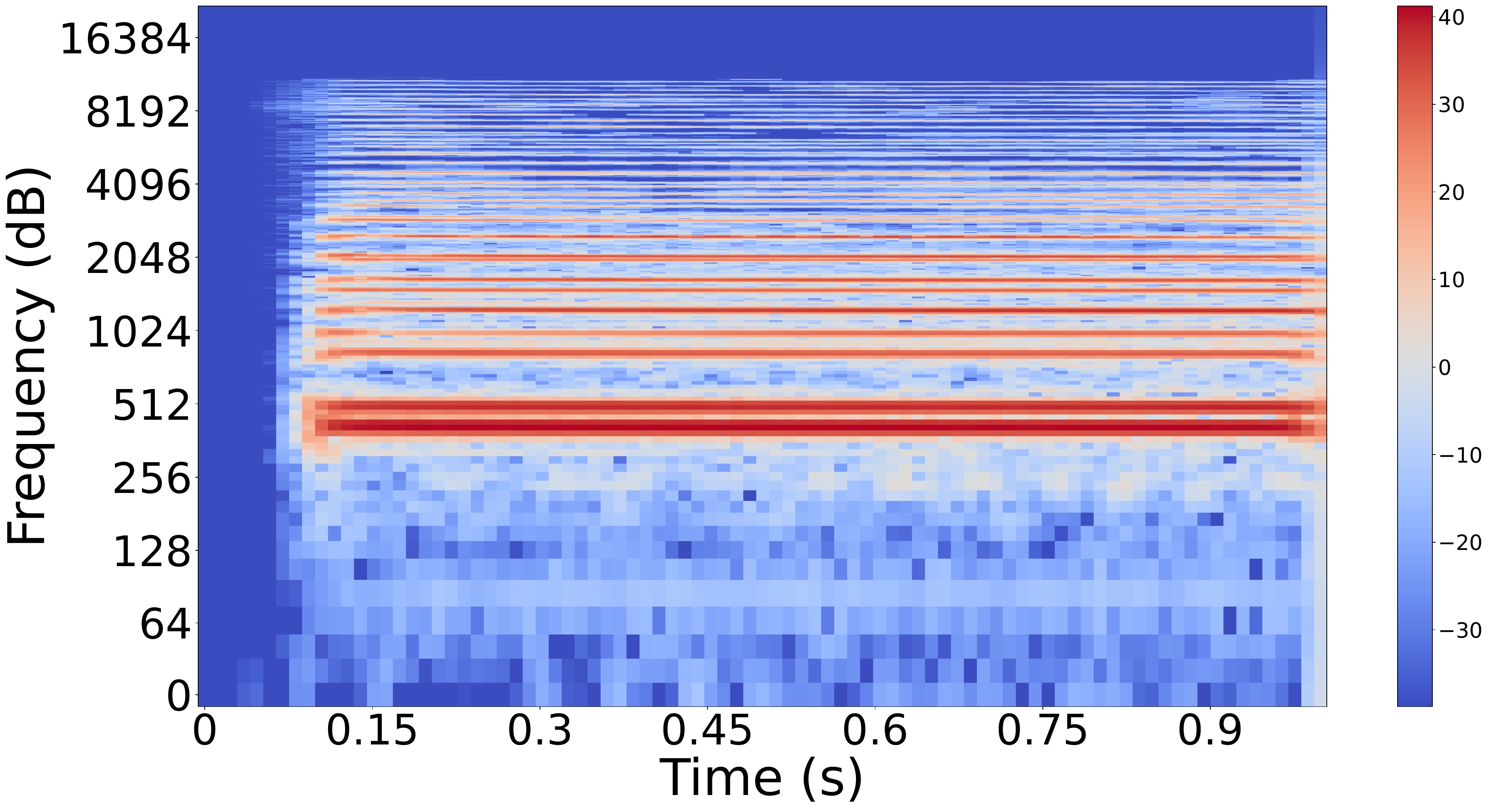}%
\label{Environmental noise}}
\hfil
\subfloat[Ultrasonic pulse]{\includegraphics[width=.32\linewidth,height=50pt]{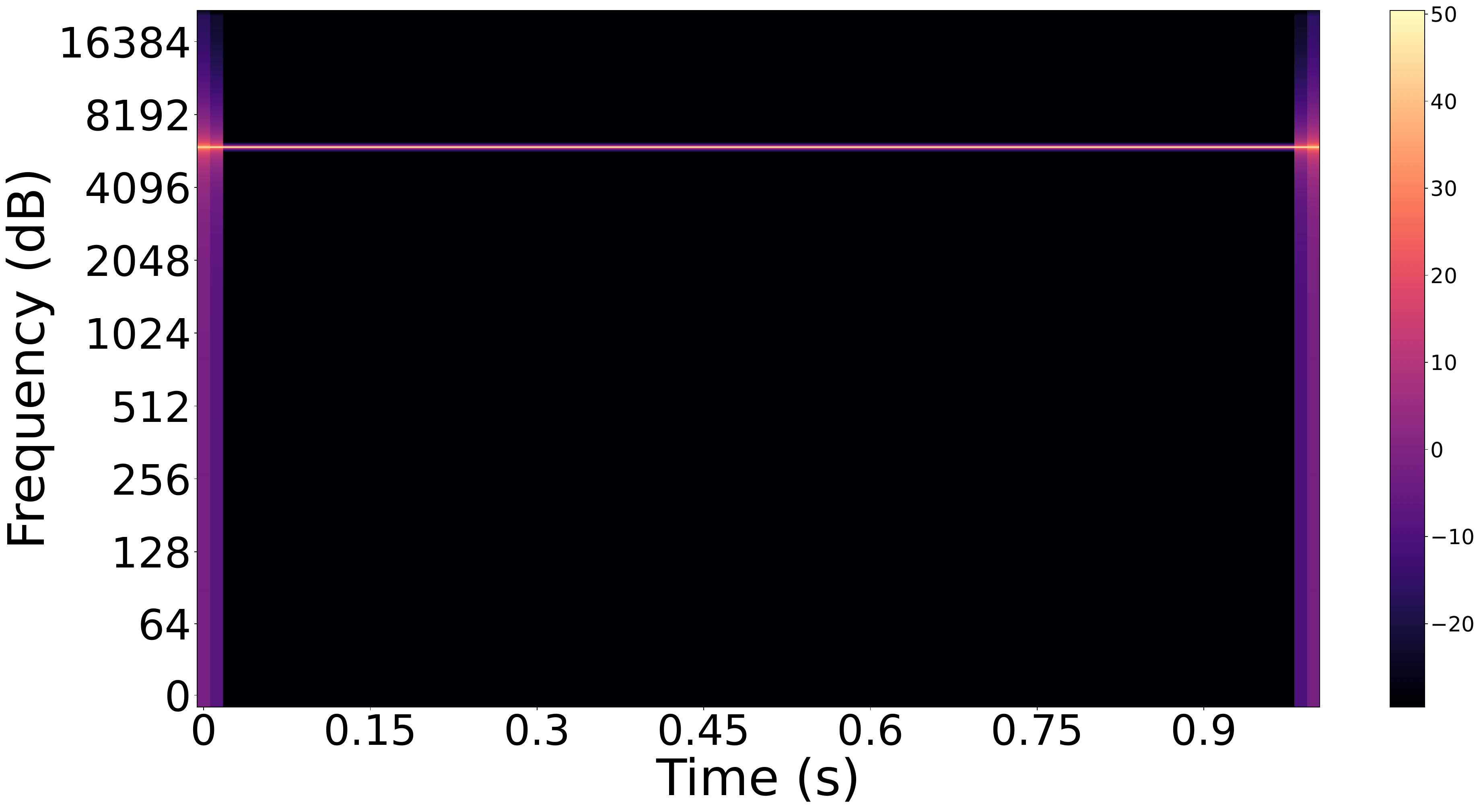}%
\label{Ultrasonic pulse}}
\hfil
\caption{Trigger pattern.}
\label{fig:4}
\end{figure}
\begin{itemize}
  \item \textbf{Random noise}: Generating poisoned speech samples by adding random noise\cite{liu2018trojaning} or perturbations\cite{tang2020embarrassingly}.
  \item \textbf{Environmental noise}:  Utilizing environmental noise (e.g., whistle sound) as triggers, characterized by the automatic activation of the backdoor\cite{liu2022opportunistic}.
  \item \textbf{Ultrasonic pulse}:  Utilizing ultrasonic pulses (6KHz in this study) as a trigger, which is inaudible to human perception\cite{koffas2022can}.
\end{itemize}
\subsubsection{Evaluation Metrics}
Backdoor attack schemes are commonly evaluated using two performance metrics:
\begin{itemize}
  \item \textbf{Benign Accuracy (BA)}: BA measures the prediction accuracy of the backdoor model on benign test samples.
  \item \textbf{Attack Success Rate (ASR)}: This metric quantifies the proportion of poisoned samples successfully directed the target label.
\end{itemize}
The objective of SpeechGuard is to maximize the reduction of ASR while maintaining BA unchanged. Additionally, to evaluate the detection capability of S-STRIP, the following metrics are introduced:
\begin{itemize}
  \item \textbf{False Rejection Rate (FRR)}: FRR represents the probability that a benign input is erroneously classified as a poisoned input by the detection system.
  \item \textbf{False Acceptance Rate (FAR)}: FAR represents the probability that a poisoned input is incorrectly classified as a benign input by the detection system.
  \item \textbf{Purification Accuracy (PA)}: PA indicates the prediction accuracy of purified poisoned samples.
\end{itemize}
In an ideal scenario, both FRR and FAR should be minimized as much as possible. However, under realistic conditions, accepting a slightly higher FRR is a common trade-off to reduce FAR.

\subsection{Performance Evaluation for Detection}
\subsubsection{Trojaned Model Performance}
We chose random noise, environmental noise and ultrasonic pulse as triggers and configured the poisoning rate to 1\% (247 poisoned samples). Table~\ref{tab_1} presents the attack performance of the backdoor model. The victim model demonstrates a prediction accuracy comparable to that of the benign model on clean samples, yet attains an attack success rate exceeding 99\%, signifying the successful implantation of the backdoor.
\begin{table}
\caption{Performance of backdoor models.}\label{tab_1}
\begin{tabular}{p{0.7cm}p{0.7cm}cccc}
\hline
\multirow{2}{*}{Dataset} & \multirow{2}{*}{Model}    & \multirow{2}{*}{ACC(\%)} & \multirow{2}{*}{Trigger} & \multicolumn{2}{c}{Trojaned model} \\ \cline{5-6} 
                         &                           &                          &                          & BA(\%)          & ASR(\%)          \\ \hline
\multirow{6}{*}{SCDv2}   & \multirow{3}{*}{2D-CNN}   & \multirow{3}{*}{92.68}   & Random noise             & 92.33           & 100              \\ \cline{4-6} 
                         &                           &                          & Environmental noise      & 92.06           & 99.86            \\ \cline{4-6} 
                         &                           &                          & Ultrasonic pulse         & 92.43           & 99.65            \\ \cline{2-6} 
                         & \multirow{3}{*}{Att-LSTM} & \multirow{3}{*}{92.80}   & Random noise             & 94.67           & 100              \\ \cline{4-6} 
                         &                           &                          & Environmental noise      & 92.88           & 99.52            \\ \cline{4-6} 
                         &                           &                          & Ultrasonic pulse         & 90.96           & 99.58            \\ \hline
\multirow{6}{*}{AMT}     & \multirow{3}{*}{2D-CNN}   & \multirow{3}{*}{99.80}   & Random noise             & 99.82           & 100              \\ \cline{4-6} 
                         &                           &                          & Environmental noise      & 99.80           & 100              \\ \cline{4-6} 
                         &                           &                          & Ultrasonic pulse         & 99.77           & 99.87            \\ \cline{2-6} 
                         & \multirow{3}{*}{Att-LSTM} & \multirow{3}{*}{99.58}   & Random noise             & 99.67           & 100              \\ \cline{4-6} 
                         &                           &                          & Environmental noise      & 99.68           & 100              \\ \cline{4-6} 
                         &                           &                          & Ultrasonic pulse         & 99.25           & 99.57            \\ \hline
\end{tabular}
\end{table}

\subsubsection{The Implementation of S-STRIP Defense}
\begin{figure}
\centering
\subfloat[AMT + 2D-CNN]{\includegraphics[width=.4\linewidth,height=80pt]{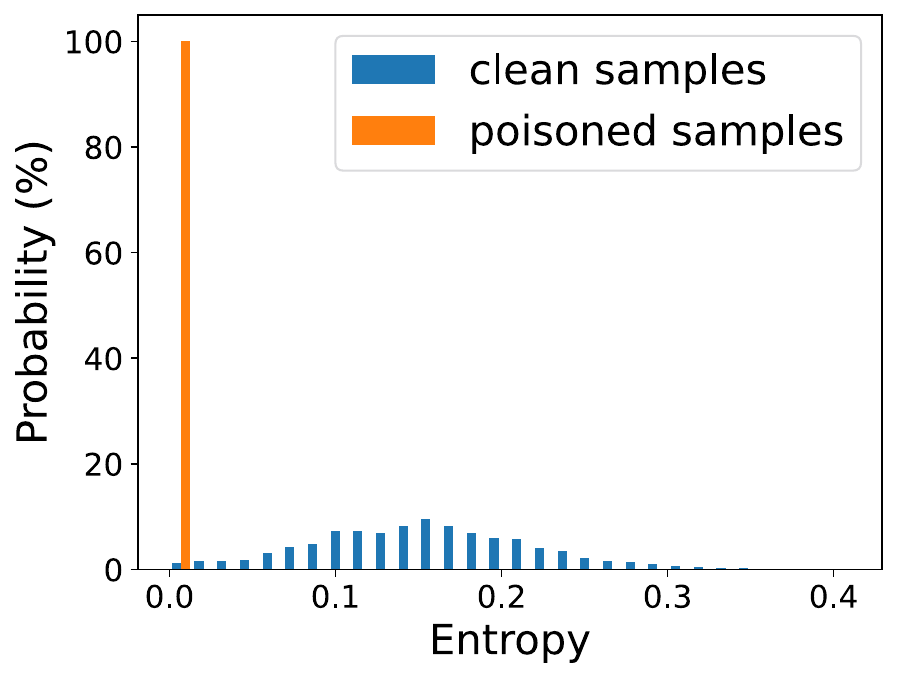}%
\label{AMT+2D-CNN}}
\hfil
\subfloat[AMT + LSTM]{\includegraphics[width=.4\linewidth,height=80pt]{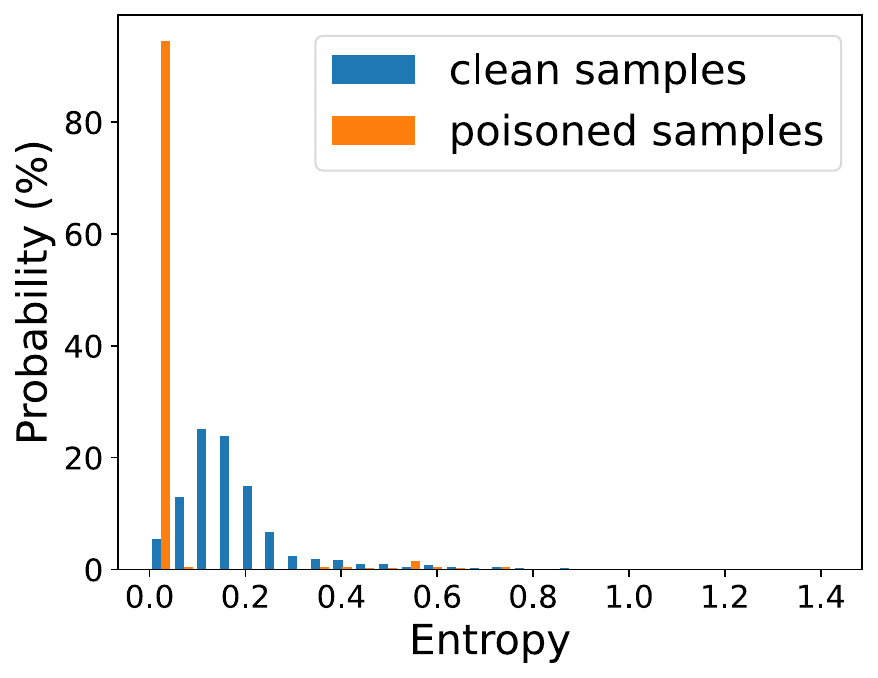}%
\label{AMT+LSTM}}
\hfil
\subfloat[SCDv2 + 2D-CNN]{\includegraphics[width=.4\linewidth,height=80pt]{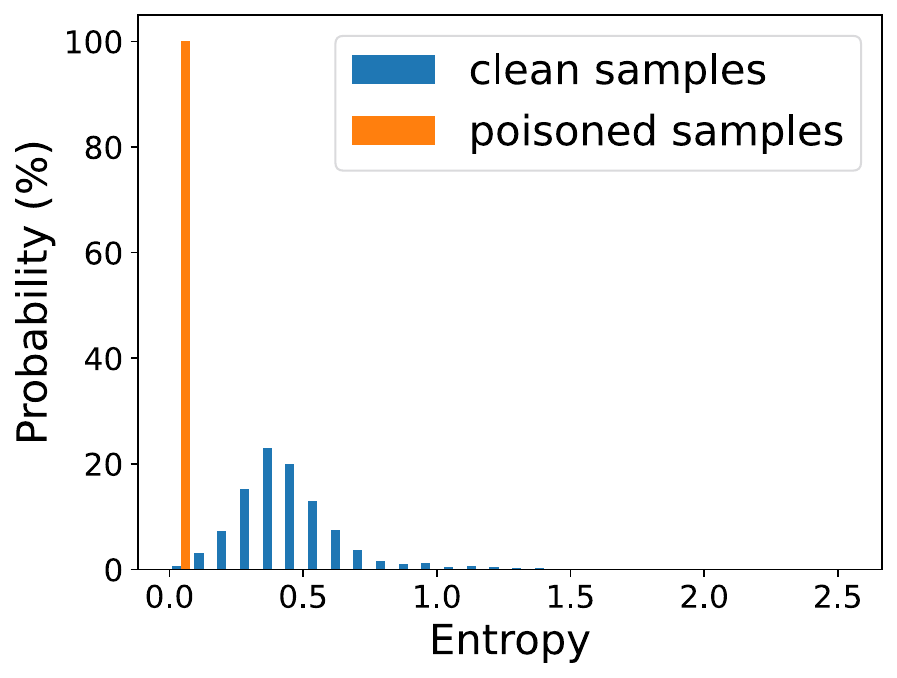}%
\label{SCDv2+2D-CNN}}
\hfil
\subfloat[SCDv2 + LSTM]{\includegraphics[width=.4\linewidth,height=80pt]{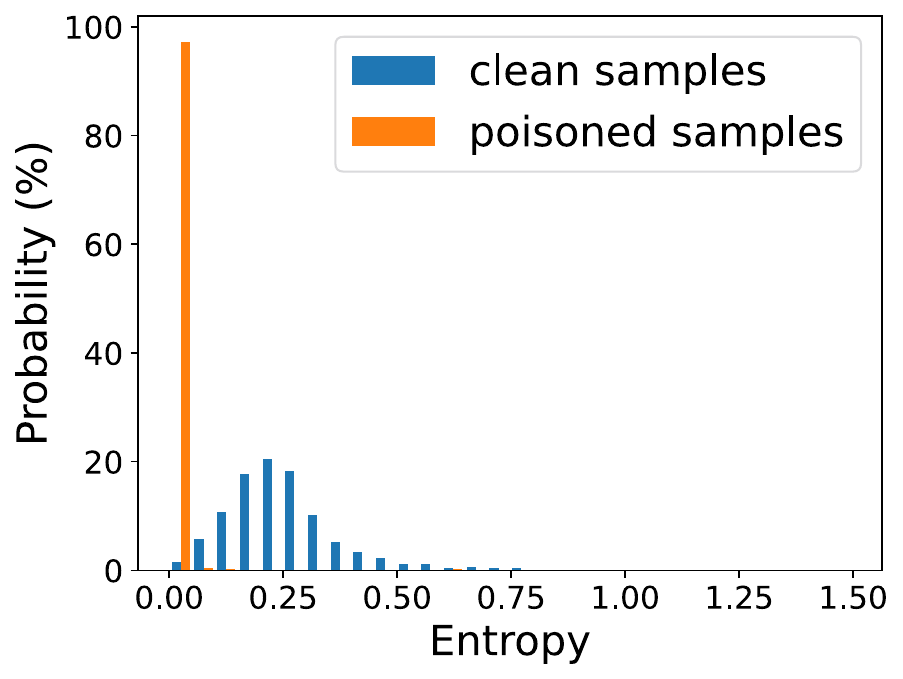}%
\label{SCDv2+LSTM}}
\caption{Entropy distribution}
\label{fig:5}
\end{figure}

We randomly selected 100 audio samples from the original dataset to serve as perturbation samples. We set SNR to 10 and determine the mixing ratio using \eqref{eq:5}. We employ 1000 raw audio samples to estimate the entropy distribution of clean samples and set the detection threshold $T$ based on the predetermined FRR. The entropy distribution of clean and poisoned samples is depicted in Fig~\ref{fig:5}.  It is evident that setting a reasonable detection threshold allows for the detection of poisoned samples.

\subsubsection{Analysis of S-STRIP Detection Performance}
The detection results are shown in Table \ref{tab_2}. In speech recognition tasks, S-STRIP demonstrates effective defense capabilities. In most instances, selecting an appropriate FRR results in a FAR below 10\%. In security-sensitive scenarios, defenders can choose a higher FRR to further reduce the FAR.
\begin{table}
\caption{S-Strip detection performance.}\label{tab_2}
\begin{tabular}{ccccc}
\hline
\multirow{2}{*}{Dataset} & \multirow{2}{*}{Model}                        & \multirow{2}{*}{Trigger}             & \multicolumn{2}{c}{Detection Metrics} \\ \cline{4-5} 
                         &                                               &                                      & FRR(\%)           & FAR(\%)           \\ \hline
\multirow{18}{*}{SCDv2}  & \multirow{9}{*}{2D-CNN}                       & \multirow{3}{*}{Random noise}        & 1                 & 0                 \\
                         &                                               &                                      & 2                 & 0                 \\
                         &                                               &                                      & 5                 & 0                 \\ \cline{3-5} 
                         &                                               & \multirow{3}{*}{Environmental noise} & 1                 & 1.50              \\
                         &                                               &                                      & 2                 & 1.10              \\
                         &                                               &                                      & 5                 & 0.75              \\ \cline{3-5} 
                         &                                               & \multirow{3}{*}{Ultrasonic pulse}    & 1                 & 0.70              \\
                         &                                               &                                      & 2                 & 0.40              \\
                         &                                               &                                      & 5                 & 0.10              \\ \cline{2-5} 
                         & \multicolumn{1}{l}{\multirow{9}{*}{Att-LSTM}} & \multirow{3}{*}{Random noise}        & 1                 & 2.90              \\
                         & \multicolumn{1}{l}{}                          &                                      & 2                 & 2.75              \\
                         & \multicolumn{1}{l}{}                          &                                      & 5                 & 2.55              \\ \cline{3-5} 
                         & \multicolumn{1}{l}{}                          & \multirow{3}{*}{Environmental noise} & 1                 & 16.10             \\
                         & \multicolumn{1}{l}{}                          &                                      & 2                 & 11.15             \\
                         & \multicolumn{1}{l}{}                          &                                      & 5                 & 6.50              \\ \cline{3-5} 
                         & \multicolumn{1}{l}{}                          & \multirow{3}{*}{Ultrasonic pulse}    & 1                 & 9.95              \\
                         & \multicolumn{1}{l}{}                          &                                      & 2                 & 7.50              \\
                         & \multicolumn{1}{l}{}                          &                                      & 5                 & 5.20              \\ \hline
\multirow{18}{*}{AMT}    & \multirow{9}{*}{2D-CNN}                       & \multirow{3}{*}{Random noise}        & 1                 & 0                 \\
                         &                                               &                                      & 2                 & 0                 \\
                         &                                               &                                      & 5                 & 0                 \\ \cline{3-5} 
                         &                                               & \multirow{3}{*}{Environmental noise} & 1                 & 0                 \\
                         &                                               &                                      & 2                 & 0                 \\
                         &                                               &                                      & 5                 & 0                 \\ \cline{3-5} 
                         &                                               & \multirow{3}{*}{Ultrasonic pulse}    & 1                 & 1.5               \\
                         &                                               &                                      & 2                 & 0.2               \\
                         &                                               &                                      & 5                 & 0.1               \\ \cline{2-5} 
                         & \multirow{9}{*}{Att-LSTM}                     & \multirow{3}{*}{Random noise}        & 1                 & 5.85              \\
                         &                                               &                                      & 2                 & 5.70              \\
                         &                                               &                                      & 5                 & 5.55              \\ \cline{3-5} 
                         &                                               & \multirow{3}{*}{Environmental noise} & 1                 & 11.40             \\
                         &                                               &                                      & 2                 & 11.05             \\
                         &                                               &                                      & 5                 & 10.00             \\ \cline{3-5} 
                         &                                               & \multirow{3}{*}{Ultrasonic pulse}    & 1                 & 7.30              \\
                         &                                               &                                      & 2                 & 6.80              \\
                         &                                               &                                      & 5                 & 6.30              \\ \hline
\end{tabular}
\end{table}
\subsection{Performance Evaluation for Purification}
\subsubsection{Autoencoder training}
The defender randomly selects 100 poisoned samples filtered through S-STRIP as inputs to the autoencoder. Time-frequency masks, typically in IBM form, are computed based on the poisoned samples and their corresponding clean samples (typically from the same label) as labels for the autoencoder. The autoencoder is subsequently trained to adaptively generate the T-F mask.
\begin{table}
\caption{Prediction results for purified poisoned samples.}\label{tab_3}
\begin{tabular}{p{0.7cm}p{1.2cm}cp{0.5cm}p{0.5cm}l}
\hline
\multirow{2}{*}{Dataset} & \multirow{2}{*}{Model}    & \multirow{2}{*}{Trigger} & \multicolumn{2}{c}{ASR(\%)} & \multirow{2}{*}{PA(\%)} \\ \cline{4-5}
                         &                           &                          & Before        & After       &                     \\ \hline
\multirow{6}{*}{SCDv2}   & \multirow{3}{*}{2D-CNN}   & Random noise             & 99.34         & 2.27        & 61.09               \\ \cline{3-6} 
                         &                           & Environmental noise      & 98.76         & 4.11        & 86.98               \\ \cline{3-6} 
                         &                           & Ultrasonic pulse         & 99.10         & 2.66        & 94.07               \\ \cline{2-6} 
                         & \multirow{3}{*}{Att-LSTM} & Random noise             & 97.95         & 1.15        & 61.21               \\ \cline{3-6} 
                         &                           & Environmental noise      & 99.68         & 1.49        & 86.37               \\ \cline{3-6} 
                         &                           & Ultrasonic pulse         & 99.28         & 1.08        & 95.18               \\ \hline
\multirow{6}{*}{AMT}     & \multirow{3}{*}{2D-CNN}   & Random noise             & 99.91         & 0.76        & 86.98               \\ \cline{3-6} 
                         &                           & Environmental noise      & 99.98         & 6.46        & 91.60               \\ \cline{3-6} 
                         &                           & Ultrasonic pulse         & 99.94         & 9.38        & 91.32               \\ \cline{2-6} 
                         & \multirow{3}{*}{Att-LSTM} & Random noise             & 98.55         & 2.91        & 72.65               \\ \cline{3-6} 
                         &                           & Environmental noise      & 99.81         & 0.35        & 93.27               \\ \cline{3-6} 
                         &                           & Ultrasonic pulse         & 99.91         & 7.14        & 93.13               \\ \hline
\end{tabular}
\end{table}
\subsubsection{Analysis of purification performance}
The prediction results for purified poisoned samples are presented in Table~\ref{tab_3}. Following purification by the autoencoder, the threat posed by the poisoned samples is significantly mitigated, with ASR decreasing by more than 90\% across all cases. An ASR below 10\% is approximately equivalent to random guessing, signifying that the trigger can no longer reliably associate with the target label. Furthermore, the prediction accuracy of purified poisoned samples remains above 60\%. Specifically, when random noise is employed as a trigger, PA drops to around 60\%. For the remaining two triggers, PA is maintained at approximately 90\%. SpeechGuard exhibits outstanding defensive capabilities against the sparsely distributed trigger. As depicted in Fig.~\ref{fig:3}, the energy distribution of random noise is more dispersed, resulting in a substantial degradation of the effective speech signal during masking. In contrast, the energy distribution of environmental noise and ultrasound is sparser, enabling the autoencoder to filter out more trigger signals while preserving more valid signal. Based on the earlier assumption that attackers prefer injecting triggers in isolated frequency bands, SpeechGuard can achieve commendable defensive performance. 

\section{DISCUSSION}
In this section, we introduce two questions to further discuss the defensive features of SpeechGuard.
\subsection{Q1:Can misdetected clean samples be accurately predicted after purification?}
When detecting poisoned samples with S-STRIP, to further minimize FAR, the defender must accept a higher FRR. In this situation, a few clean samples may be erroneously identified as poisoned and subjected to the subsequent purification step. Hence, we aim to investigate whether purified clean samples can still maintain a high prediction accuracy. To simulate this scenario, we directly input 1000 clean samples into the autoencoder for purification and then fed the purified samples into the backdoor model for prediction. The experimental results are presented in Table \ref{tab_4}. Excluding random noise as a trigger, prediction accuracy of misdetected clean samples does not drop by more than 10\%. This demonstrates that our purification approach accurately identifies and eliminates trigger signals. Even in cases of false positives in the first-stage detection scheme, there is no significant loss in accuracy, further ensuring the robustness of the SpeechGuard defense solution.
\begin{table}[]
\caption{Prediction accuracy for purified clean samples..}\label{tab_4}
\begin{tabular}{p{0.7cm}p{1.2cm}cp{0.5cm}p{0.5cm}l}
\hline
\multirow{2}{*}{Dataset} & \multirow{2}{*}{Model}    & \multirow{2}{*}{Trigger} & \multicolumn{3}{c}{BA(\%)}                    \\ \cline{4-6} 
                         &                           &                          & before & after & \multicolumn{1}{l}{decline} \\ \hline
\multirow{6}{*}{SCDv2}   & \multirow{3}{*}{2D-CNN}   & Random noise             & 96.90  & 60.44 & 36.46                        \\ \cline{3-6} 
                         &                           & Environmental noise      & 96.98  & 86.39 & 10.59                        \\ \cline{3-6} 
                         &                           & Ultrasonic pulse         & 97.15  & 93.90 & 3.25                         \\ \cline{2-6} 
                         & \multirow{3}{*}{Att-LSTM} & Random noise             & 92.17  & 60.42 & 31.75                        \\ \cline{3-6} 
                         &                           & Environmental noise      & 93.40  & 85.64 & 7.76                         \\ \cline{3-6} 
                         &                           & Ultrasonic pulse         & 95.68  & 95.05 & 0.63                         \\ \hline
\multirow{6}{*}{AMT}     & \multirow{3}{*}{2D-CNN}   & Random noise             & 99.97  & 85.87 & 14.10                         \\ \cline{3-6} 
                         &                           & Environmental noise      & 99.95  & 90.90 & 9.05                         \\ \cline{3-6} 
                         &                           & Ultrasonic pulse         & 99.92  & 90.87 & 9.05                         \\ \cline{2-6} 
                         & \multirow{3}{*}{Att-LSTM} & Random noise             & 99.83  & 71.45 & 28.38                        \\ \cline{3-6} 
                         &                           & Environmental noise      & 98.77  & 92.87 & 5.90                          \\ \cline{3-6} 
                         &                           & Ultrasonic pulse         & 99.80  & 92.80 & 7.00                            \\ \hline
\end{tabular}
\end{table}
\subsection{Q2:Why is IBM chosen as the target masking?}
\begin{figure}
\centering
\includegraphics[width=250pt]{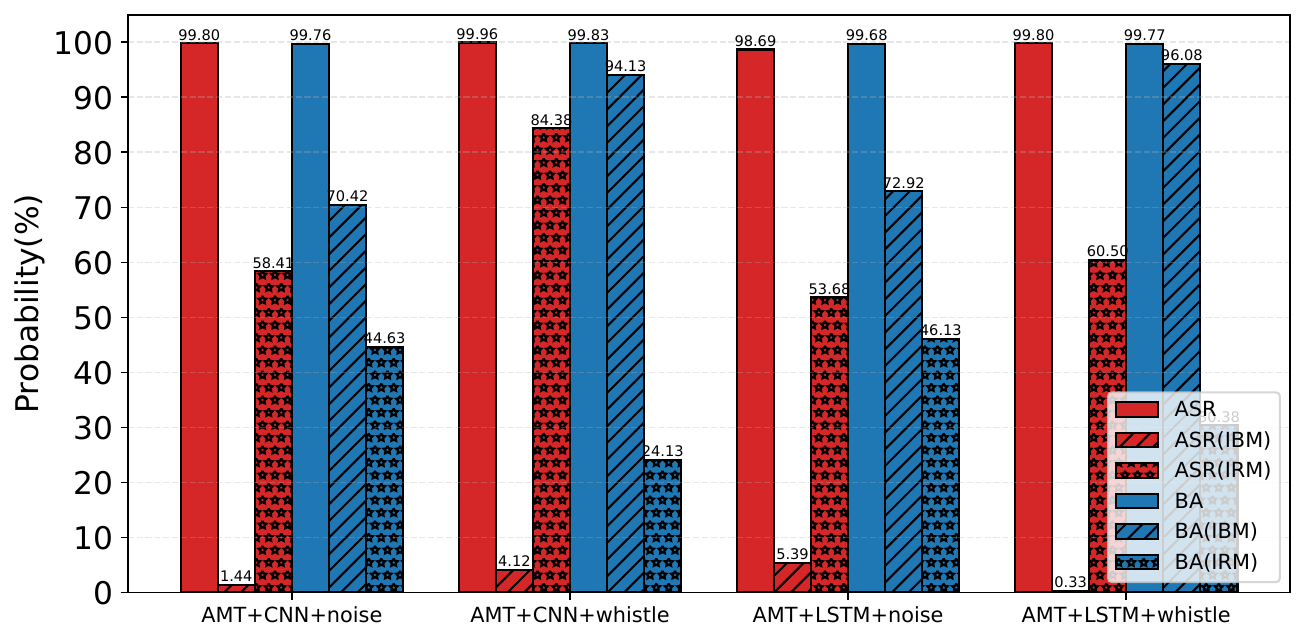}
\caption{Comparison of purification performance between IBM and IRM.}
\label{fig:6}
\end{figure}

SpeechGuard employs IBM as the default target masking. Here, we explore the possibility of better alternatives, such as IRM, formulated as \eqref{eq:12}: 
\begin{equation}
\mathbf{M}_{IRM}(t,f)=\left( \frac{|X(t,f)|^2}{|X(t,f)|^2+|N(t,f)|^2} \right)^{\beta},
\label{eq:12}
\end{equation}
We evaluated the efficacy of IBM and IRM masking on four backdoor models, and the results are presented in Fig~\ref{fig:6}. IBM demonstrates a distinct advantage over IRM, achieving a more substantial reduction in attack threat while preserving higher prediction accuracy. As a soft decision, IRM retains more valid speech signals but fails to thoroughly suppress trigger signal. Interestingly, preserving additional speech signals does not enhance prediction accuracy, as backdoor models tend to prioritize the residual trigger signals. Consequently, we recommend adopting IBM as the target masking to more effectively eliminate trigger signals.

\section{Conclusion}
In this paper, we introduce SpeechGuard, an online backdoor defense solution comprising a two-stage pipeline for poisoned sample detection and purification. In the detection stage, we propose an enhanced S-STRIP detection method derived from STRIP, incorporating perturbations with a predetermined blend ratio to adapt it for speech recognition tasks. In the purification stage, we leverage the sparse distribution of audio signals in the time-frequency domain, proposing the utilization of masking learned from an autoencoder to suppress trigger signals. Extensive experiments on two popular datasets and models validate the effectiveness of SpeechGuard. Significantly, SpeechGuard focuses on speech recognition tasks, providing a comprehensive defense pipeline from detection to purification, thereby addressing existing gaps in current research.

\vspace{-2\baselineskip}
\enlargethispage{5\baselineskip}
\bibliographystyle{unsrt}  
\bibliography{references} 

\end{document}